\documentstyle[epsfig,array,aps]{revtex}
\begin{document}
\draft
\title{Decoherence and fidelity in ion traps with fluctuating trap parameters}
\author{S. Schneider and G.~J.~Milburn}
\address{Centre for Laser Science, Department of Physics,
The University of Queensland, St. Lucia,
QLD 4072, Australia}
\date{\today}
\maketitle

\begin{abstract}
We consider two different kinds of fluctuations in an ion trap potential:
external fluctuating electrical fields, which cause
statistical movement
(``wobbling'') of the ion relative to the center of the trap,
and fluctuations of the spring constant, which are due to fluctuations
of the ac-component of the potential applied in the Paul trap for ions.
We write down master
equations for both cases and, averaging out the noise,
obtain expressions for the heating of the ion.
We compare our results to previous results for far-off
resonance optical traps and heating in ion traps.  The
effect of fluctuating external electrical fields
for a quantum gate
operation (controlled-NOT) is determined and the fidelity for that
operation derived.
\end{abstract}
\date{today}

\section{Introduction}
The ability to engineer and control pure quantum states of trapped ions is
driving a number of
new technologies including time and frequency measurements, new
measurement schemes and
quantum logic manipulations for quantum computation\cite{LANL,NIST,NIST2}. The
key requirement for engineering quantum states is the
necessity to remove, or at least control, all sources of noise and
uncertainty.  Laser cooling in particular enables
uncertainty about the initial vibrational state to be removed by cooling
the ions to the collective vibrational ground
state.  From that point pure states may be prepared using highly stabilized
laser pulses. Despite these achievements, however,
noise cannot be entirely eliminated. Residual laser intensity and phase
fluctuations in the pulses that are used
to shape the quantum states must be taken into
account\cite{Schneider1998,Murao1998} as well as noise in the trapping
parameters.
In
this paper we consider various sources of
noise in the trap itself
and determine their effect on cold trapped ions and the ability to
perform reversible logical operations.

Noise is of course the origin of decoherence, the process which limits the
ability to maintain pure quantum states. However,
we need to be careful in making the connection between noise and
decoherence. From a fundamental perspective the dynamics of
the ions is always unitary and reversible, even in the presence of noise,
but by definition the dynamics of a noisy
quantity is uncontrollable and often unknown. Thus the precise unitary
dynamics varies from
one run of the experiment to the next and  the exact motion of the state in
Hilbert space may not be known or even precisely
predictable. In the course of the experiment we do not have
precise control over, or knowledge of, the unitary transformations in state
engineering, and thus we cannot be sure we have
reached the desired state in Hilbert space.  In the case of
quantum computation such a result is manifest as an
error. Faced with describing such a system we can simply average over the
noise, which in practical terms means we combine the
data from many experiments all performed with different realizations of the
noisy control parameters. Alternatively we can
give the sample space of error states in each run, together with their
probability of occurrence.  In this paper we
give both descriptions particularly for the
case of fluctuations of the center (equilibrium
point) of the trap potential.

In an ion trap an inability to precisely control the motion leads to an
unwanted excitation of the vibrational state of the ion,
that is 'heating'. The
main source of this heating appears to be due to the ambient fluctuating
electrical fields in the trap.  There are now a number
of experiments that have measured this
heating\cite{NIST,King98,Monroe95a,Diedrich89}.  Recently
James\cite{James98}
has shown that a simple theory of this source of heating can be given in
terms of a harmonic oscillator subject to a
fluctuating classical driving field.  Our model for this source of heating
is similar, although our theoretical description is
a little different.

The paper is organized as follows: In section II we first derive the
heating rate of the ion due to fluctuating electrical fields.
Section III is devoted to the effects of fluctuations
in the spring constant of the trap potential on the heating.
In the fourth section we
look at the effects of fluctuating electrical fields on gate operations. As
a specific example we investigate the effects on the so far
experimentally realized  controlled-NOT gate (NIST gate) \cite{Monroe1995b}.
We conclude with a discussion of our results.

\section{Heating due to fluctuating electrical fields}

We want to model the effects of fluctuating electrical fields with the
same formalism used in \cite{Schneider1998} and compare our results to those
obtained by Savard \em et al.\em {} \cite{savard}. There they derive
the heating rate for a far-off resonance optical trap due to
fluctuations in the location of the trap center which are caused
by laser-beam-pointing noise. We apply the formalism here to
an ion in a Paul trap. In those traps
additional electrical fields
cause a replacement of the center of the trap, to which the ion
adjusts automatically by finding the minimum of the potential.
We assume that those electrical fields which
cause this replacement have got an additional white noise component,
to which the ion cannot adjust.

The Hamiltonian in this case is
\begin{equation}
H = \frac{p^2}{2m} + \frac{1}{2} m \omega^2 x^2 + \xi(t) x \,\, ,
\label{lin-fluct}
\end{equation}
where $x$ and $p$ are the position and momentum operator of the ion in
the trap, $m$ is the mass of the ion, and $\omega$ is the trap frequency.
The term $\xi(t)$ describes the fluctuating electrical force due to
fluctuating fields which we
assume to be due to a white noise process, i.e.
\begin{equation}
\xi(t) dt = \sqrt{\gamma} d W(t) \,\, ,
\end{equation}
with $dW(t)$ a Wiener process and the parameter $\gamma$ scales the noise.
Since we are dealing with a white noise process, we define a stochastic
Schr\"odinger equation in the Ito formalism \cite{dyrting1996}
\begin{equation}
d \rho = - \frac{i}{\hbar} [H_0, \rho] dt - \frac{i}{\hbar} \sqrt{\gamma}
[x, \rho] dW(t) - \frac{\gamma}{2 \hbar^2} [x, [x, \rho]] \,\, ,
\end{equation}
where
\begin{equation}
H_0 = \frac{p^2}{2m} + \frac{1}{2} m \omega^2 x^2 \,\, .
\end{equation}
For a single run with a known (or monitored) behaviour of the noise in
time, the above equation gives the evolution of the system density operator
conditioned on the entire history of the noise process. Since we are
not interested in the effects of the fluctuations in some short time
limit or just for one run of the experiment, we follow \cite{Schneider1998}
and average over the noise to get the master equation for the average
density operator
\begin{equation}
\frac{d \tilde{\rho}}{dt} = - \frac{i}{\hbar} [H_0, \tilde{\rho}]
- \frac{\gamma}{2 \hbar^2} [x, [x, \tilde{\rho}]] \,\, .
\end{equation}
This equation has a 'high frequency' limit which may be relevant in
experiments for which the time scale is much longer than
the period of the trap. For example if the trapped ions form a quantum
logic gate the time over which the gate
operation is imposed may be much greater than the trap period (which is
typically of the order of $10^{-6}$s). In that case we
can transform to a frame rotating at the trap frequency and time average
the rapidly rotating terms to give a master equation
in the form
\begin{equation}
\frac{d\rho}{dt}=\frac{\gamma}{2\hbar m\omega}\left (a\rho
a^\dagger+a^\dagger\rho a-\frac{1}{2}(a^\dagger
a\rho+aa^\dagger\rho+\rho a^\dagger a+\rho aa^\dagger)\right ) \,\, . 
\label{time-av}
\end{equation}

To calculate the mean energy
\begin{equation}
\langle H_0 \rangle(t) = \frac{\langle p^2\rangle}{2m}
+ \frac{1}{2} m \omega^2 \langle x^2 \rangle \,\, ,
\end{equation}
we look at the time derivative of $\langle x^2 \rangle$ and
$\langle p^2 \rangle$ and get the two equations
\begin{eqnarray}
\dot{\langle x^2\rangle} & = & - \frac{1}{m} \langle xp + px \rangle \\
\dot{\langle p^2\rangle} & = & m \omega^2 \langle xp + px \rangle +
\gamma \,\,,
\end{eqnarray}
and thus
\begin{equation}
\frac{d \langle H_0 \rangle}{dt} = \frac{1}{2m} \gamma \,\, .
\end{equation}
So the mean energy is
\begin{equation}
\langle H_0 \rangle (t) = \frac{1}{2 m} \gamma t + \langle H_0
\rangle_{(t=0)} \,\, .
\label{wobble_hot}
\end{equation}
(The same result obtains with the time averaged master equation as well).
This result is equivalent to that derived by
Savard \em et al. \em \cite{savard} and by James \cite{James98}
in the limit of white
noise and with appropriate changes of notation.

To make the comparison with the results of Wineland \em et al. \em \cite{NIST} 
we write
Eq.~(\ref{wobble_hot}) in terms of the mean vibrational quantum number
$\bar{n}$ as
\begin{equation}
\frac{d\bar{n}}{dt}=\frac{1}{t^*} \,\, ,
\end{equation}
where the time constant for decoherence is then given by
\begin{equation}
t^*=\frac{2\hbar m\omega}{\gamma} \,\, .
\end{equation}
The fluctuating linear potentials are caused by fluctuating electric
fields, $E(t)$, on the trap electrodes thus we expect
that the fluctuating term in Eq.~(\ref{lin-fluct}) is given by $\xi(t)=qE(t)$.
If the fluctuations in the electric field are
treated as white noise the spectral density of these fluctuations (near the
trap frequency) is independent of frequency. If we
take $E(t)dt=E_0dW(t)$ then we have the equivalence $\gamma=q^2E_0^2$ with
the spectral density of fluctuations in the field
given by
\begin{eqnarray}
S_E(\omega) & = & 4\int_0^\infty \overline{E(t+\tau)E(t)}d\tau\\
  & = & 2E_0^2 \,\, .
\end{eqnarray}
Thus the decoherence time becomes
\begin{equation}
t^*=\frac{4\hbar m\omega}{q^2S_E(\omega)} \,\, ,
\end{equation}
which is the form quoted in reference \cite{NIST}.

\section{Noise in the spring constant}

In an ion trap a suitable combination of ac and dc electric fields are used
to create
an approximate harmonic potential in three spatial dimensions for one or
more ions\cite{Ghosh1995}. In the case of more than
one ion, the coulomb force couples the motion of the ions and the
collective normal mode coordinates undergo harmonic motion.
Laser cooling enables the ion, or the collective mode of many ions, to be
prepared in or near the ground state of the system.
Noise in the spring constant in  due to
fluctuations in the ac-component and dc-components of the applied
potential. In the linear trap of reference\cite{NIST}
with the long axis oriented along the z-axis, and with micromotion ignored,
the periodic trap potential at the trap centre is
given by
\begin{equation}
\Phi\approx \frac{1}{2}m\omega_z^2+\frac{1}{2}m\omega_r^2(x^2+y^2) \,\, ,
\end{equation}
where the harmonic force in the $z$-direction is formed by a static potential
and in the $x-y$ direction it is formed by
an ac potential at frequency $\Omega_T$ and amplitude $V_0$. The resonance
frequency is then given by
$ \omega_r=qV_0/(2^{1/2}\Omega_TmR^2)$, 
where $m$ and $q$ are the ion mas and charge and $R$ is the distance from the
$z$-axis to the surface of the linear electrodes.
Clearly fluctuations in either the dc or ac component will lead to
fluctuations in the spring constants of the trap, although
in practice ac fluctuations are more significant.

In far-off resonance
optical traps, the restoring force is provoced by the induced optical
dipole force of an applied laser. Typically the atom is
confined at the node of a standing wave of a laser tuned above the atomic
resonance (blue detuning). The atom them sees a
mechanical potential proportional to the intensity of the laser. If the
intensity is quadratic near the node, a linear
restoring force will be produced. In this case
fluctuations in the applied laser intensity lead to fluctuations in the trap
frequency\cite{savard}.

The Hamiltonian for a particle moving in a harmonic potential with
fluctuating spring constant is
\begin{equation}
H = \frac{p^2}{2m} + \frac{1}{2} m \omega^2 \left(1 + \epsilon(t)\right)
x^2 \,\, ,
\end{equation}
where again we assume white noise,
\begin{equation}
\epsilon(t)dt = \sqrt{\Gamma} d W (t) \,\, ,
\end{equation}
and the noise this time is scaled with the
parameter $\Gamma$.
To make things easier, we introduce dimensionless coordinates
\begin{eqnarray}
x & \rightarrow & \left(\frac{2 \hbar}{m \omega}\right)^{-1/2} x
= X \\
p & \rightarrow & \left( 2 \hbar m \omega\right)^{-1/2} p
= P \,\, ,
\end{eqnarray}
so that the commutation relation in these new coordinates now reads
\begin{equation}
[X, P] = \frac{i}{2} \,\, .
\end{equation}
Thus our Hamiltonian takes on the form
\begin{eqnarray}
H & = & \hbar \omega \left( P^2 + X^2\right) + \epsilon(t)
\hbar \omega X^2 \nonumber \\
  & = & H_0 + \epsilon(t)\hbar \omega X^2 \,\, .
\end{eqnarray}

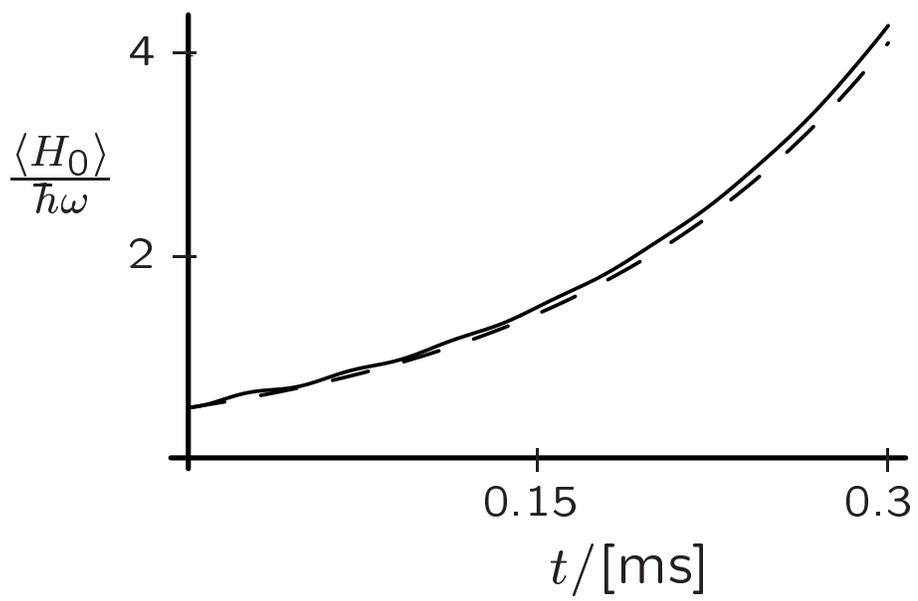
\begin{figure}
\caption{Plot of the exact (solid line) and approximated (dashed line)
solution for the increase of the phonon number in time. This plot is for a
vibrational frequency $\omega = 11.2 (2\pi)$ kHz and for $\Gamma \omega/2
= 0.1$. We choose the initial values to be $\langle X^2\rangle_{t=0} =
\langle P^2 \rangle_{t=0} = (1/2) \langle XP + PX \rangle_{t=0} =
1/4$, so that $\langle H_0 \rangle_{t=0} /(\hbar \omega) = 1/2$.}
\label{fig1}
\end{figure}

Again we are not interested in a specific noise history, but rather in the
overall effect of the fluctuations. So we average out the noise and the
corresponding master equation reads
\begin{equation}
\frac{d \tilde{\rho}}{d t} = - \frac{i}{\hbar} [H_0, \tilde{\rho}] -
\frac{\Gamma}{2}
 \omega^2 [X^2, [X^2, \tilde{\rho}]] \,\, .
\end{equation}
We want to determine the change of the mean energy with time.
The mean energy is given by the expression
\begin{equation}
\langle H_0 \rangle (t) = \hbar \omega \left( \langle P^2 \rangle (t)
+ \langle X^2 \rangle (t) \right) \,\, .
\end{equation}
We wish to get the expressions for $ \langle P^2 \rangle (t)$ and
$\langle X^2 \rangle (t)$. To do this, we can first derive a system of
first order differential equations for
$\langle P^2 \rangle (t)$, $ \langle X^2 \rangle (t)$ and
$\langle XP + PX \rangle (t) /2$:
\begin{equation}
\frac{d}{dt} \left(
\begin{array}{c}
\langle X^2 \rangle\\
\langle P^2 \rangle\\
\frac{1}{2} \langle XP + PX \rangle
\end{array} \right)
= A
 \left(
\begin{array}{c}
\langle X^2 \rangle\\
\langle P^2 \rangle\\
\frac{1}{2} \langle XP + PX \rangle
\end{array} \right) \,\, ,
\label{syse5}
\end{equation}
where
\begin{equation}
A = \left( \begin{array}{ccc}
0 & 0 & 2 \omega \\
 \Gamma \omega^2 & 0 & -2 \omega \\
- \omega & \omega & 0 \end{array} \right) \,\, .
\label{matrix}
\end{equation}
So we have to solve this system of differential equations
to get the solution for $\langle X^2 \rangle
(t)$ and $\langle P^2 \rangle (t)$.
The exact solution is
\begin{eqnarray}
\langle H_0\rangle (t) &  = & \hbar \omega \Bigg\{ \bigg[
\exp(\frac{2 (D^2 - 1)
\omega}{\sqrt{3} D} t) \frac{(2 - D^2 + 2 D^4) (1 + D^2 + D^4)}
{9 D^2 ( 1 - D^2 + D^4)} \nonumber \\
&  & + {} \exp( - \frac{(D^2 -1 ) \omega}{\sqrt{3} D} t)
\Big(- \frac{2 (D^2 - 1)^4}{9 D^2 ( 1 - D^2 + D^4)}
\cos(- \frac{(1 + D^2)\omega}{D} t) \nonumber \\
&  &
+ {} \frac{2 (1 - D^2)(1 + 2 D^2 - D^{4} + D^6 + D^8)}
{3 \sqrt{3} D^2 (1 + D^6)}
\sin(- \frac{(1 + D^2)\omega}{D} t) \Big) \bigg] \langle X^2 \rangle_{t=0}
\nonumber \\ &  &
+ {} \bigg[ \exp( \frac{2 (D^2 - 1) \omega}{\sqrt{3} D} t)
\frac{ 1 + D^2 + D^4}{3 (1 - D^2 + D^4)}  \nonumber \\
&  & +{}
 \exp( - \frac{(D^2 -1 ) \omega}{\sqrt{3} D} t)
\Big( \frac{2 (D^2 -1)^2}{3 (1 - D^2 + D^4)}
\cos(- \frac{(1 + D^2)\omega}{D} t)
\nonumber \\
&  & +{} \frac{2 D^2 (D^2 -1)}
{\sqrt{3} (1 + D^2) (1 + D^2 + D^4)}
\sin(- \frac{(1 + D^2)\omega}{D} t)\bigg] \langle P^2 \rangle_{t=0}
\nonumber \\
&  & {} + \bigg[\exp(\frac{2 (D^2 - 1) \omega}{\sqrt{3} D} t)
\frac{2 (D^6 - 1)}{3 \sqrt{3} D (1 - D^2 + D^4)} \nonumber \\
&  & + {} \exp( - \frac{(D^2 -1 ) \omega}{\sqrt{3} D} t)
\Big( - \frac{2 (D^6 - 1)}{3 \sqrt{3} D (1 - D^2 + D^4)}
\cos(- \frac{(1 + D^2)\omega}{D} t) \nonumber \\
&  & + {}\frac{2 (D^2 -1)^2 (1 + D^2 + D^4)}{3 D
 (1 + D^6)} \sin(- \frac{(1 + D^2)\omega}{D} t)\Big)
\bigg]\frac{1}{2} \langle XP + PX \rangle_{t=0} \Bigg\} \,\, ,
\end{eqnarray}
where
\begin{equation}
D = \sqrt[3]{\frac{3 \sqrt{3}}{4} \frac{\Gamma \omega}{2} + \sqrt{1 +
\frac{27}{16} \left( \frac{\Gamma \omega}{2} \right)^2}} \,\, .
\end{equation}
To simplify things, we  assume that the noise is a small effect
compared to the free dynamics, i.e.
\begin{equation}
\frac{\Gamma}{2} \omega \ll 1 \,\, .
\end{equation}
Using this approximation we get
\begin{equation}
\langle H_0 \rangle (t) = \hbar \omega \exp \left[ \frac{\Gamma
\omega^2}{2} t \right] \left(
\langle P^2 \rangle_{(t=0)} + \langle X^2 \rangle_{(t=0)} \right) \,\, .
\end{equation}
Again this result is equivalent to the one given by Savard {\it et al.}
\cite{savard} in the limit of white noise.
In Fig.~\ref{fig1}  both the exact (solid line)
and the approximated (dashed line) solution are plotted
for the vibrational frequency $\omega =  11.2  (2 \pi)$ kHz and the value
of $\Gamma \omega / 2$ is 0.1 in this plot. We choose the initial values
to be $\langle X^2\rangle_{t=0} = \langle P^2 \rangle_{t=0} =
(1/2) \langle XP + PX \rangle_{t=0} = 1/4$, so that $\langle H_0\rangle_{t=0}
/\hbar \omega = 1/2$.

%

\section{Effects of fluctuating external electrical
fields during quantum gate operations}

We have investigated the overall effects of
the fluctuations in the spring constant and the
position in ion traps. It is of interest to look at the effects of those
fluctuations
on gate operations used for quantum computation \cite{Cirac1995}.
The gate operations are performed by shining an additional laser, with
a specific frequency and for a well determined time, on a
two-level transition in an ion.
This laser causes an interaction between the internal electronic states
of the ion and the CM motion of the ion or of all the ions if there is
more than one ion in the trap.

We will
denote the electronic qubit as $|g\rangle$, $|e\rangle$ for the ground
state and the excited state
respectively. The coding is such that the ground state is logical $0$ while
the excited state is logical $1$.
The
vibrational state will be denoted by the energy eigenstates $|0\rangle_v$,
$|1\rangle_v$ which are the ground state and the first
excited state respectively.
%

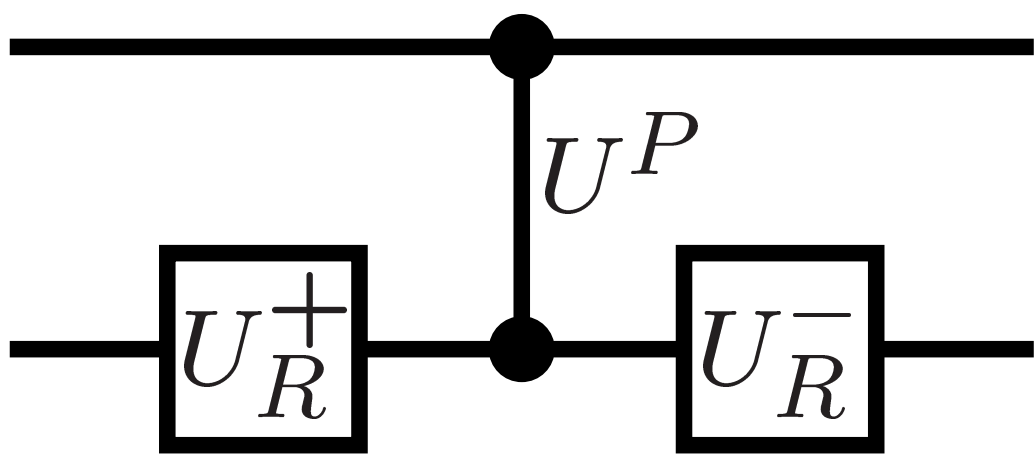
\begin{figure}
\caption{Schmeatic representation of a controlled-NOT gate: The controlled
phase shift $U_P$ is sandwiched between the two single qubit rotations
$U^+_R$ and $U^-_R$.}
\label{fig2}
\end{figure}

A controlled-NOT gate can be broken down into the circuit shown in figure
\ref{fig2}, where a controlled
phase shift is sandwiched between two $\pi/2$ pulses with different
phases on the target qubit,
which in this case is the electronic qubit.
The $\pi/2$ pulses produce rotations of the electronic qubit:
\begin{eqnarray}
U^+_R & : & \left\{
\begin{array}{l}
|0\rangle\rightarrow
1/\sqrt{2}\left(|0\rangle-|1\rangle\right)\\
|1\rangle\rightarrow
1/\sqrt{2}\left(|0\rangle+|1\rangle\right)
\end{array}\right. \\
U^-_R & : & \left\{
\begin{array}{l}
|0\rangle_{j}\rightarrow
1/\sqrt{2}\left(|0\rangle+|1\rangle\right)\\
|1\rangle_{j}\rightarrow
-1/\sqrt{2}\left(|0\rangle-|1\rangle\right)
\end{array}\right. \,\, .
\end{eqnarray}
The controlled phase shift $U_P$ acts to produce
a $\pi$ phase shift only if both
the electronic qubit and the vibrational qubit is
in the logical state  $|1\rangle_L$. The total transformation from input to
output is then given by
\begin{equation}
|\Psi_{out}\rangle=U_R^- U_P U_R^+ |\Psi_{in}\rangle \,\, .
\label{Ugate}
\end{equation}

We take the most general input state to the controlled-NOT gate as
\begin{equation}
|\Psi_{in}\rangle= (\alpha |g\rangle + \beta |e\rangle) \otimes (\delta
|0 \rangle_v + \epsilon |1 \rangle_v)
\label{initstate}
\end{equation}
with $\alpha$, $\beta$, $\gamma$ and $\delta$ being complex amplitudes
satisfying
\begin{equation}
|\alpha|^2 + |\beta|^2 = |\delta|^2 + |\epsilon|^2 = 1 \,\, .
\end{equation}
The first $\pi/2$ pulse $U^+_R$ acting on the target (electronic) qubit, then
produces the state;
\begin{equation}
|\Psi_{1} \rangle
= \frac{\alpha + \beta}{\sqrt{2}} \delta |g \rangle |0 \rangle_v
+ \frac{\beta - \alpha}{\sqrt{2}} \delta |e \rangle |0 \rangle_v
+ \frac{\alpha + \beta}{\sqrt{2}} \epsilon |g \rangle |1 \rangle_v
+ \frac{\beta - \alpha}{\sqrt{2}} \epsilon |e \rangle |1 \rangle_v \,\, .
\label{Hinitstate}
\end{equation}
where the subscript $1$ indicates that this is the state after the first
rotation.
The controlled phase shift operation just changes the sign of the last term.
Then the final rotation $U^-_R$ takes this state to the output state
\begin{equation}
|\Psi_{out} \rangle
= \alpha \delta |g\rangle|0\rangle_v + \beta \delta |e \rangle |0\rangle_v +
\alpha \epsilon |e\rangle |1\rangle_v + \beta \epsilon |g \rangle |1\rangle_v
\,\, .
\label{ideal}
\end{equation}
If the vibrational qubit is subject to noise in the trap center position,
i.e.~there are fluctuating electrical fields,
the gate will not operate as required. The
noise does not effect the rotations, which only involve the
electronic qubit (except for heating up the vibarional state during those
rotations), however, it will disrupt the
controlled phase shift operation which couples the electronic and
vibrational systems.
We first make the usual transformation
into the interaction picture  with
\begin{equation}
U_0=\exp \left[ \frac{i}{\hbar} H_0t\right] \,\, ,
\label{trafo}
\end{equation}
where
\begin{equation}
H_0 = \hbar \omega a^\dagger a + \hbar \omega_A \sigma_+
\sigma_-  \,\, ,
\end{equation}
to give
\begin{equation}
H_I = H_G-\hbar\lambda \xi(t) \left( a^\dagger
e^{i \omega t} + a e^{-i \omega t} \right) \,\, ,
\label{Hint}
\end{equation}
where $H_G$ is the interaction picture Hamiltonian causing the gate
operation and  where
\begin{equation}
\lambda=\left (2\hbar m \omega \right)^{-1/2} \,\, .
\end{equation}

To include the noise we need to calculate the effective controlled phase
shift operation including the noise term
over the time of the gate operation.
The delta-correlated nature of white noise enables a simple approach in
which noise terms can be separated and treated
perturbatively while the gate interaction is treated to all orders.  To
enable this approximation we first transform to an
interaction picture defined by
\begin{equation}
|\Psi(t)\rangle=\exp\left (-\frac{i}{\hbar}H_G t\right )|\Psi_I(t)\rangle
\,\, .
\end{equation}
The total time evolution for the controlled phase-shift gate over the gate
time $T$ is given by
$|\Psi^\prime_2\rangle=U^\prime_P|\Psi_1\rangle$ with
\begin{eqnarray}
U^\prime_P & = & \exp\left (-\frac{i}{\hbar}H_G T\right ){\cal T}
\left(\exp\left
[-\frac{i}{\hbar}\int_0^Tdt^\prime
H_{noise}(t^\prime)\right]\right) \nonumber \\
& = & U_PU_N[\xi(t)] \,\, ,
\label{timeorder}
\end{eqnarray}
where
\begin{equation}
H_{noise}(t)=\exp\left (\frac{i}{\hbar}H_G t\right )\left (-\hbar\lambda
\xi(t) \left( a^\dagger
e^{i \omega t} + a e^{-i \omega t} \right)\right )\exp\left
(-\frac{i}{\hbar}H_G t\right )
\label{aver}
\end{equation}
and
where ${\cal T}$ is the time ordering operator and $T$ the time required
for the gate operation in the absence of noise.
We have indicated the functional dependence on a particular realization of
the noise by $[\xi(t)]$.
The time-ordered evolution operator appearing as the second factor in
Eq.~(\ref{timeorder}) may be treated perturbatively
in the stochastic amplitude, $\xi(t)$, by carrying the Dyson expansion to
second order.

There are two ways to view the effects of the noise. One way is at the
level of a single realization of a gate operation.
This view enables us to see what the error states will be in the presence
of noise. The second way is to determine the result
of a gate operation by ensemble averaging the noise. This view enables us
to give an average fidelity for the gate operation
in the presence of the noise. The two views will be referred to as a
quantum trajectory picture and an ensemble average picture,
respectively. In the quantum trajectory picture the output state is a
functional of the particular noise history, $\xi(t)$,
over the gate operation time $T$. The output state is then a pure state of
the form
\begin{equation}
|\Psi^\prime[\xi(t)]\rangle=U_R^- U_P U_N[\xi(t)]|\Psi_1\rangle \,\, ,
\end{equation}
where, as earlier, $|\Psi_1\rangle$ is the state after the first
rotation of the electronic qubit. To find the likely
error states the noise operator can now be expanded in powers of the noise
amplitude.

In the ensemble picture we need to average over all possible realizations
of the noise over the gate time $T$. The output
state is now a mixed state given by
\begin{equation}
\rho^\prime_{out}=U_R^- U_P \rho_1^\prime U_P U_R^-
\label{Ngate}
\end{equation}
and
\begin{equation}
\rho_1^\prime=\int
U_N([\xi])|\Psi_1\rangle\langle\Psi_1|U_N^\dagger([\xi])
P[\xi]d[\xi]
\,\, ,
\end{equation}
where $P[\xi]d[\xi]$ is the probability functional for each
noise realization.  To calculate $\rho_1^\prime$  we expand the unitary
operator $U_N[\xi(t)]$ to second order in
the noise and then average over the classical stochastic variables.  The
evolution of the density operator over the gate time
$T$ is then given by a Dyson-expansion\cite{louisell} which to second
order is
\begin{equation}
\rho_1^\prime  \cong  \rho_1 + \frac{1}{i \hbar} \int_0^T [H_{noise}
(t_1), \rho_1] dt_1
 +  \left(\frac{1}{i \hbar}\right)^2 \int_0^T dt_1 \int_0^{t_1} dt_2
[H_{noise}(t_1), [H_{noise}(t_2), \rho_1]] \,\, .
\label{dyson}
\end{equation}
Since $E(dW(t))=0$, the average over the
second term vanishes and we only have to calculate  the average over the
third term.
This is done by noting that\cite{Gardiner}
\begin{equation}
\left\langle\int_0^t G(t^\prime) dW(t^\prime)
\int_0^{t^\prime} G(t^{\prime\prime})
d W(t^{\prime\prime})\right\rangle = \int_0^t G(t^\prime)^2 dt^\prime
\,\, .
\end{equation}

We can quantify the effect of noise on the average through the fidelity
defined by
\begin{equation}
F(\gamma)= \langle \Psi_{out}|\rho^\prime_{out}|\Psi_{out}\rangle \,\, ,
\label{fidel1}
\end{equation}
where $|\Psi_{out}\rangle$ is the output state for a noiseless gate
operation while $\rho^\prime_{out}$ is the output state of
the gate averaged over all realizations of the noise.  The
fidelity is the probability that the system is in the desired state
and will
depend on the noise correlation strength $\gamma$. Substituting
Eqs.~(\ref{Ugate}, \ref{Ngate}) we find that
\begin{equation}
F(\gamma)=\langle\Psi_1|\rho^\prime_1| \Psi_1\rangle\,\, .
\label{fidel2}
\end{equation}
To proceed we take two examples for realizing a controlled-NOT gate between
the vibrational state and the internal state for one ion in the trap.

\subsection{Mutual phase-shift gate}

In the discussion of the perfect controlled-NOT gate we saw that
the essential two-qubit
operation is a controlled phase shift.
We will discuss two different ways by which this can be done. The first way
involves a
mutual conditional phase shift of the vibrational and electronic degrees of
freedom. This operator commutes with the
vibrational quantum number.  The second way involves an auxiliary
electronic level and is used in the NIST scheme to produce
a controlled-NOT gate\cite{Monroe1995b}.  We first discuss the mutual phase
shift gate as this is simpler.

The mutual phase shift gate is defined by
the unitary transformation
\begin{equation}
U_P=\exp\left (-i\pi a^\dagger a\otimes|e\rangle\langle e|\right ) \,\, .
\label{mutualphase}
\end{equation}
To include the noise we need to calculate the effective controlled
phase shift operation including the noise term
over the time of the gate operation,
where the gate Hamiltonian is
\begin{equation}
H_G= \hbar\kappa a^\dagger a \otimes|e\rangle\langle e|
\end{equation}
and with the gate operation time such that $\kappa T=\pi$, where $\kappa$
is a constant. An interaction
of this kind can be produced by a carrier frequency
excitation of the ion \cite{Vogel}.

The total time evolution is then given by Eq.~(\ref{timeorder}) with
\begin{equation}
H_{noise}=-\hbar\lambda\xi(t)\left (a\exp\left [-i\kappa t|e\rangle\langle
e|-i\omega t\right ]+ h.c.\right ) \,\, ,
\end{equation}
where $h.c.$ indicates the hermitian conjugate of the preceeding term. The
effect of the noise is determined by expanding
the second time ordered factor in Eq.~(\ref{timeorder}) to second order, thus
\begin{equation}
U_N  = 1+i(a\nu(T)+a^\dagger\nu^\dagger(T))
-\frac{1}{2}\int_0^T dt \left
\{a\frac{d\nu(t)}{dt}+a^\dagger\frac{d\nu^\dagger(t)}{dt}\right \}
\left \{a\nu(t)+a^\dagger\nu^\dagger(t)\right \} \,\, ,
\end{equation}
where the operators $\nu$ are defined in the following way:
\begin{equation}
\nu(t)  =  \lambda\sqrt{\gamma}\int_0^t
dW(t^\prime)e^{-i(\omega+\kappa |e\rangle\langle e|) t^\prime} \,\, .
\end{equation}

We first determine the likely error states in the quantum trajectory
picture. The desired gate operation,
Eq.~(\ref{mutualphase}),
commutes with the phonon number operator $a^\dagger a$ and cannot change
the vibrational quantum number,
and thus the vibrational states always remains in the logical basis of
$|0\rangle$, $|1\rangle$. The noise factor $U_N$,
however, is linear in $a$ and $a^\dagger$ and thus does change the phonon
number.  Keeping terms to second order in the
stochastic amplitude means that, in a single realization of the gate, the
effect of the noise is to `leak'  coherence into the
vibrational states $|2\rangle_v$, $|3\rangle_v$ as well as changing the
weighting of the vibrational qubit basis states
$|g\rangle$, $|e\rangle$.  For example, in the quantum trajectory picture, a
single realization of the noise takes the
state Eq.~(\ref{Hinitstate}) to the state
\begin{eqnarray}
|\Psi^\prime_2\rangle &  = &  e^{-i\pi a^\dagger a|e\rangle\langle e|}
\Big(|\Psi_{1} \rangle
 + {} \frac{\alpha + \beta}{\sqrt{2}} \delta\nu_g^*|g \rangle |1\rangle_v
+  \frac{\beta - \alpha}{\sqrt{2}} \delta\nu_e^*|e\rangle |1\rangle_v
\nonumber\\ &  & + {}
\left. \frac{\alpha + \beta}{\sqrt{2}} \epsilon
(\nu_g|g \rangle |0 \rangle_v+\sqrt{2}\nu_g^*|g \rangle |2
\rangle_v)
  +  \frac{\beta - \alpha}{\sqrt{2}} \epsilon
\nu_e|e \rangle |0 \rangle_v+\sqrt{2}\nu_e^*|e \rangle
|2 \rangle_v)+\ldots\right ) \,\, ,
\end{eqnarray}
where we have suppressed the second order terms, and the subscript $2$
indicates that this is the state after the controlled
phase shift gate operation, and the prime indicates a state corrupted by
noise. The first term is the correct output state and
all subsequent terms correspond to an error. The error terms are multiplied
by random variables defined by
\begin{eqnarray}
\nu_g & = & \lambda\sqrt{\gamma}\int_0^TdW(t)e^{-i\omega t}\\
\nu_e & = & \lambda\sqrt{\gamma}\int_0^TdW(t)e^{-i(\omega+\kappa) t}\,\, .
\end{eqnarray}
If the gate time $T$ is large compared to the vibrational frequency, these
random variables have zero mean and the
correlation functions are\cite{Gardiner}
\begin{eqnarray}
E(\nu_g\nu_g)=E(\nu_e\nu_e) & = & 0\\
E(\nu_g^*\nu_g)=E(\nu_e^*\nu_e) & = & \lambda^2\gamma T\\
E(\nu_g^*\nu_e) & = & \frac{2i\lambda^2\gamma T}{\pi} \,\, .
\end{eqnarray}

As mentioned above we can calculate the fidelity of the gate operation,
Eq.~(\ref{fidel1}, \ref{fidel2}). The result up to first
in $\gamma$ is
\begin{eqnarray}
F(\Gamma_\kappa) & = &1 -2 \Gamma_\kappa (1 + 2 |\epsilon|^2) +
\Gamma_\kappa
(1 - |\epsilon|^2) |\epsilon|^2
\left[ |\beta - \alpha|^4 \left( 1 + \frac{\kappa}{ \pi (\omega + \kappa)}
\sin(\frac{\omega \pi}{\kappa}) \cos(\frac{\omega \pi}{\kappa} + 2 \Delta)
\right) \right.
\nonumber \\
&  &
+ {} \left. |\alpha + \beta|^4  \left( 1 + \frac{\kappa }{\pi \omega}
\sin( \frac{\omega \pi}{\kappa} ) \cos ( \frac{\omega \pi}{\kappa} +
2 \Delta \right)  - |\beta - \alpha |^2 |\alpha + \beta|^2 \frac{ 4 \kappa}
{\pi (2 \omega + \kappa)} \cos( \frac{\omega \pi}{\kappa}) \sin( \frac{\omega
\pi}{\kappa} + 2 \Delta) \right]
\end{eqnarray}
where
\begin{equation}
\Gamma_\kappa = \frac{\pi \gamma}{\hbar m \omega \kappa} \,\,
\end{equation}
is the now dimensionless noise parameter and $\Delta$ is the phase
difference between $\delta$ and $\epsilon$
\begin{equation}
\Delta = \phi_\delta - \phi_\epsilon \,\, ,
\label{phasediff}
\end{equation}
where we denote the phase of $\delta$ ($\epsilon$) by $\phi_\delta$
($\phi_\epsilon$), respectively.

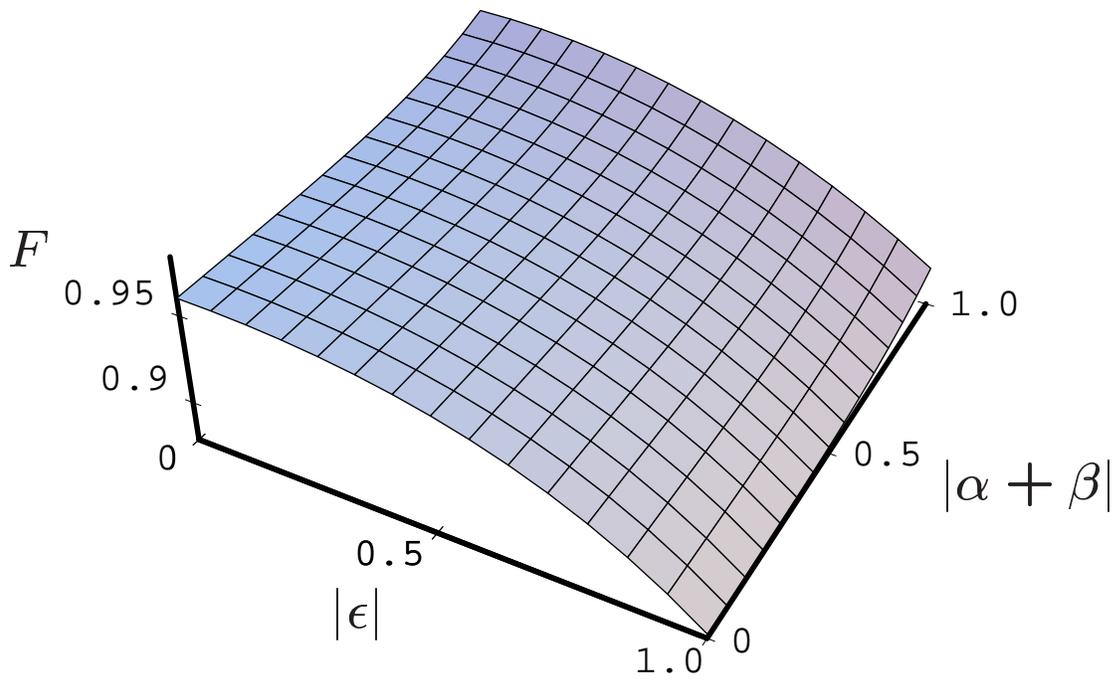
\begin{figure}
\caption{Plot of the dependence of
$F$ on the initial state we want to perform the gate on.
The chosen parameters for the plot are $\omega = 11.0 (2 \pi)$ MHz,
$\kappa = 1.0 $ MHz, $\Delta = 0$, and $\Gamma_\kappa = 0.02$.}
\label{fig3}
\end{figure}

As expected for an expansion up to second order in $\sqrt{\gamma}$ the
fidelity depends linearly on $\gamma$, since we have averaged out the
first order term. The inverse dependance on the nonlinear coupling constant
$\kappa$ is easy to understand. If this parameter
is large, the gate time can be made very short in which case the noise has
less time to act and produce an error. Thus the
fidelity should approach one as $\kappa$ becomes large.  Note that in
general the fidelity depends on the initial state. The dependence of 
$F$ on the initial state is plotted in Figure \ref{fig3}. The plot 
parameters are $\omega = 11.0 (2 \pi)$ MHz, $ \Omega \eta = 1.0 (2 \pi)$ 
kHz, $\Delta = 0$, and $ \Gamma_a = 0.02$.

\subsection{NIST Gate}
The Hamiltonian required to describe the controlled phase shift operation
in the NIST gate \cite{Monroe1995b} is
\begin{equation}
H_G = \hbar \frac{\Omega_a \eta}{2}
\left(|e\rangle \langle \mbox{aux}|  a^\dagger + |\mbox{aux}\rangle
\langle  e | a \right)
\end{equation}
where
$\eta$ is the Lamb-Dicke-parameter, and $a$ and $a^\dagger$ are the
creation and
annihilation operators for the quantized CM motion of the ion.
This Hamiltonian is used to describe a $2 \pi$ blue-sideband pulse between
the excited state of the electronic qubit (logical
1) and an auxiliary state $|\mbox{aux}\rangle$. The rotations of
the electronic states are carried out by $\pi/2$
pulses that are assumed to act without noise. This assumption is
reasonable since the effective Rabi frequencies for those transitions
are higher than those for the blue sideband pulses, thus leading to
much shorter pulse durations.

As before we separate the pure gate operation from the noise by
transforming to an interaction picture through the gate
Hamiltonian.  
We then determine the average fidelity for the gate operation by averaging
over the noise.

Calculating $\rho^\prime_1$ using Eq.~(\ref{dyson}) and in particular
Eq.~(\ref{aver}) is a rather tedious, but straightforward process.
The fidelity rate turns out to be
\begin{eqnarray}
F(\Gamma_a) & = & 1 - 2 \Gamma_a  \left[ 1 + |\epsilon|^2 \right]
- \Gamma_a |\epsilon|^2  |\alpha + \beta|^2
\nonumber \\
&  & + {}
\Gamma_a |\epsilon|^2 (1 - |\epsilon|^2) |\beta - \alpha|^4
 \sin \left( \frac{4 \pi \omega}{\Omega \eta}\right)
\cos\left( \frac{4 \pi \omega}{\Omega \eta} + 2 \Delta \right)
\left[\frac{\Omega \eta}{8 \pi \omega } - \frac{\Omega \eta}{16 \pi (\Omega
\eta -
2 \omega)} + \frac{\Omega \eta}{16 \pi (\Omega \eta + 2 \omega)} \right]
\nonumber \\
&  & + {}
 \Gamma_a |\epsilon|^2 (1 - |\epsilon|^2) |\alpha + \beta|^4
\left[ 1 + \frac{\Omega \eta}{2 \pi \omega} \sin\left( \frac{4 \pi \omega}
{\Omega \eta} \right) \cos\left( \frac{4 \pi \omega}{\Omega \eta} +
2 \Delta\right) \right]
\nonumber \\
&  & + {}
 \Gamma_a |\epsilon|^2 (1 - |\epsilon|^2) |\alpha + \beta|^2
|\beta - \alpha|^2 \sin \left( \frac{4 \pi \omega}{\Omega \eta} \right)
\cos\left( \frac{4 \pi \omega}{\Omega \eta} + 2 \Delta\right)
\left[ \frac{\Omega \eta}{\pi (\Omega  \eta - 4 \omega)} +
\frac{\Omega \eta }{\pi (\Omega \eta + 4 \omega)} \right] \,\, ,
\end{eqnarray}
where
\begin{equation}
\Gamma_a = \frac{4 \pi \gamma}{\hbar m \omega \Omega \eta}
\end{equation}
is now the new dimensionless noise parameter and $\Delta$
is again the phase difference between the two vibrational states as
defined above in Eq.~(\ref{phasediff}).

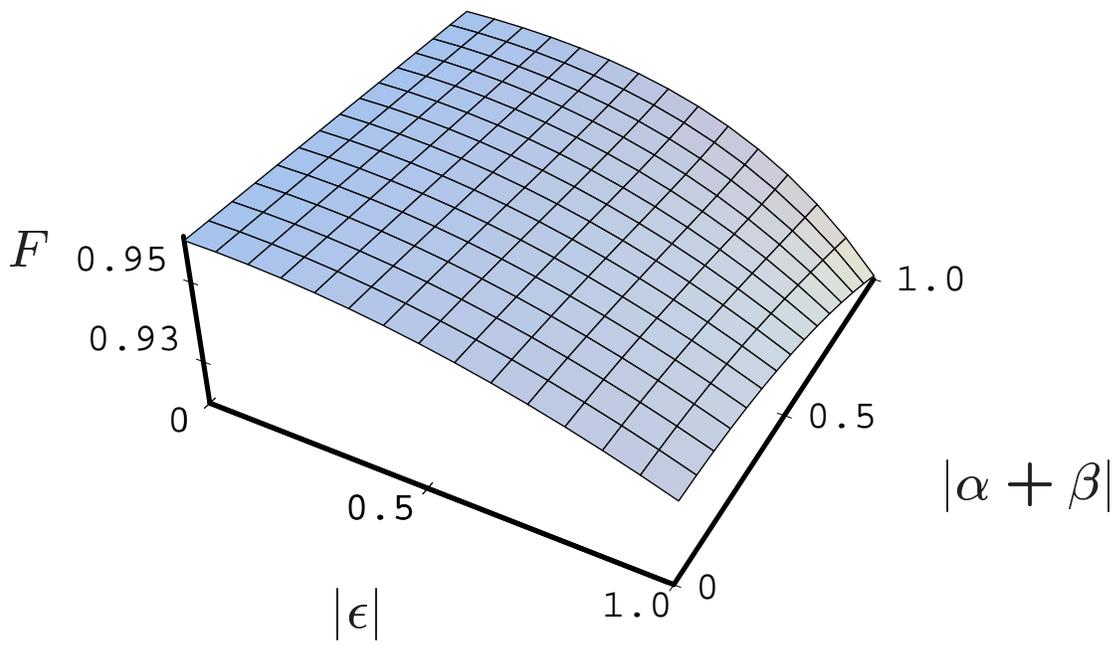
\begin{figure}
\caption{Plot of the dependence of
$F$ on the initial state we want to perform the gate on for the NIST gate.
The chosen parameters for the plot are $\omega = 11.0 (2 \pi)$ MHz,
$\Omega \eta = 12.0 (2 \pi)$ kHz, $\Delta = 0$, and $\Gamma_a = 0.02$.}
\label{fig4}
\end{figure}

Again the fidelity depends linearly on $\Gamma_a$. The dependence of
$F$ on the initial state we want to perform the gate on is plotted
in Fig.~\ref{fig4}. The chosen paramters for the plot 
are $\omega = 11.0 (2 \pi)$ MHz,
$\Omega \eta = 12.0 (2 \pi)$ kHz, $\Delta = 0$, and $\Gamma_a = 0.02$.

\section{Discussion}
We have determined the heating rates due to a fluctuating
trap potential and due to fluctuating electrical fields  for
the mena motional energy of an ion in a rf Paul trap.
The potential use of ion traps as simple quantum computers in
view we have calculated the effects of fluctuating electrical
fields (considered one of the major sources of noise at the moment)
during a controlled-NOT gate operation. We derived fidelities for
two different ways of performing the required conditional phase shift
needed for those gates. This analysis is particularly useful for
the application of ion traps to quantum computation. It gives an
estimate on how strong those fluctuating fields can be to still
perform a computation with a certain accuracy. 

Taking the current heating rate of the COM mode for the NIST ion 
trap \cite{King98}, which is about 19 phonons per ms and assuming
that these heating rate is due to fluctuating electrical fields 
(the reasons for those heating rates are not clear yet, so we just 
assume at this stage until experimentalists will come up with 
more elaborate data), we get a rough estimate for $\Gamma_a \approx 
0.02$. With that value we get fidelities above 90 \% for one gate 
operation, which certainly needs improvement to allow for more than one gate
operation.

\section*{Acknowledgments}
While completing parts of this work, one of us (SS) was visiting
Los Alamos National Laboratory. She would like to thank both the T-6 group
and the P-23 group for their hospitality. She is grateful for useful
discussions and comments from Raymond Laflamme and Daniel James.
SS gratefully acknowledges financial
support from an University of Queensland Postgraduate Research Scholarship,
from the Centre for Laser Science and from the Fellowship Fund - Branch
of AFUW Qld.~Inc.
GJM acknowledges the support of the
Australian Research Council.


\newpage
\section*{Figures}

\begin{references}
\bibitem{LANL}R.J.~Hughes, D.F.V.~James, J.J.~Gomez, M.S.~Gulley,
M.H.~Holzscheiter, P.G.~Kwiat, S.K.~Lamoreaux, C.G.~Peterson,
V.D.~Sandberg, M.M.~Schauer, C.M.~Simmons, C.E.~Thorburn,
D.~Tupa, P.Z.~Wang, and A.G.~White, Fortschritte der Physik {\bf 46},
329 (1998).

\bibitem{NIST}D.J.~Wineland, C.~Monroe, W.M.~Itano, D.~Leibfried,
B.E.~King, and D.M.~Meekhof, "Experimental issues in coherent
quantum-state manipulation of trapped atomic ions",
Journal of Research of the National
Institute of Standards and Technology {\bf 103},259 (1998).

\bibitem{NIST2} D.J.~Wineland, C.~Monroe, W.M.~Itano, B.E.~King,
D.~Leibfried, D.M.~Meekhof, C.~Myatt, and C.~Wood, Fortschritte der Physik
{\bf 46}, 363 (1998).

\bibitem{Schneider1998}S.~Schneider and G.J.~Milburn, Phys. Rev. A {\bf 57},
3748 (1998).

\bibitem{Murao1998}M.~Murao and P.L.~Knight, Phys. Rev. A {\bf 58}, 663
(1998).

\bibitem{King98}B.E.~King, C.J.~Myatt, Q.A.~Turchette, D.~Leibfried,
W.M.~Itano, C.~Monroe, and D.J.~Wineland, Phys. Rev. Lett. (in press) 1998,
and e-print quant-ph/9803023.

\bibitem{Monroe95a}C.~Monroe, D.M.~Meekhof, B.E.~King, W.M.~Itano,
and D.J.~Wineland, Phys. Rev. Lett. {\bf 75}, 4011
(1995).

\bibitem{Diedrich89}F.~Diedrich {\it et al.},
Phys. Rev. Lett. {\bf 42} 67 (1989).

\bibitem{James98}D.F.V.~James, Phys. Rev. Lett. {\bf 81} 317 (1998).

\bibitem{savard} T.A.~Savard, K.M.~O'Hara, and J.E.~Thomas, Phys. Rev. A.
{\bf 56}, R1095 (1997).

\bibitem{Monroe1995b} C.~Monroe, D.M.~Meekhof, B.E.~King, W.M.~Itano,
and D.J.~Wineland, Phys. Rev. Lett. {\bf 75}, 4714 (1995).

\bibitem{dyrting1996}S.~Dyrting and G.J.~Milburn, Quantum Semiclass. Opt.
{\bf 8}, 541 (1996).

\bibitem{louisell}W.H.~Louisell, {\it Quantum Statistical Properties
of Radiation}, (John Wiley \& Sons, New York, 1990).

\bibitem{Cirac1995} J.I.~Cirac and P.~Zoller, Phys. Rev. Lett. {\bf 74},
4094 (1995).

\bibitem{Vogel}C.~D'Helon and G.J.~Milburn, Phys. Rev. A {\bf 54}, 5141 (1996).

\bibitem{Gardiner} C.W.~Gardiner, {\it Handbook of Stochastic Methods},
(Springer, Berlin, 1985).

\bibitem{Ghosh1995} P.K.~Ghosh, {\it Ion Traps}, (Clarendon Press, Oxford,
1995).
\end{references}
\end{document}